# Two-Dimensional 8-State Potts Model on Random Lattices: A Monte Carlo Study

*Wolfhard Janke*[1] *and Ramon Villanova*[2]

[1] Institut für Physik, Johannes Gutenberg-Universität Mainz
Staudinger Weg 7, 55099 Mainz, Germany

[2] Matematiques Aplicades, Universitat Pompeu Fabra,
La Rambla 32, 08002 Barcelona, Spain

**Abstract**

We use two-dimensional Poissonian random lattices of Voronoi/Delaunay type to study the effect of quenched coordination number randomness on the nature of the phase transition in the eight-state Potts model, which is of first order on regular lattices. ¿From extensive Monte Carlo simulations we obtain strong evidence that the phase transition remains first order for this type of quenched randomness. Our result is in striking contrast to a recent Monte Carlo study of quenched bond randomness for which the order of the phase transition changes from first to second order.

# 1 Introduction

For systems exhibiting a continuous phase transition in the pure case it is well known that the influence of quenched random disorder can modify the critical behaviour, leading to new universality classes or even eliminating the phase transition altogether [1]. Also for systems undergoing a first-order phase transition the effect can be very dramatic. Phenomenological renormalization-group arguments suggest that the addition of quenched randomness can smoothen the transition completely and induce instead a continuous phase transition [2]. For a certain type of quenched bond-disorder in the two-dimensional $q$-state Potts model with $q = 8$ this prediction has recently been confirmed by extensive Monte Carlo simulations [3]. While the pure model is exactly known to exhibit a quite strong first-order phase transition [4], the simulations with quenched bond-disorder gave clear evidence for a continuous phase transition. From careful finite-size scaling analyses the phase transition was identified to belong to the Ising model universality class.

In this note we report Monte Carlo simulations of the same model subject to a different kind of quenched disorder. Instead of using a square lattice with uniform coordination number (= 4) and drawing the coupling strengths randomly from two different values as in Ref.[3], we consider Poissonian random lattices where the coordination numbers vary locally between 3 and $\infty$ and the coupling strengths are uniform. The random lattices are constructed according to the Voronoi/Delaunay prescription [5] for toroidal topology, i.e., with periodic boundary conditions.

# 2 Model and simulation

We used the standard definition of the $q$-state Potts model,

$$Z_{\text{Potts}} = \sum_{\{\sigma_i\}} e^{-\beta E}; E = -\sum_{\langle ij \rangle} \delta_{\sigma_i \sigma_j}; \sigma_i = 1, \ldots, q, \tag{1}$$

where $\beta = J/k_B T$ is the inverse temperature in natural units, and $\langle ij \rangle$ denotes the nearest-neighbor bonds of random lattices with $V = 250$, 500, 750, 1000, 2000, and 3000 sites. For each lattice size we generated 20 independent replica and performed long simulations of the 8-state model near the transition point at $\hat{\beta} = 0.826$, 0.830, 0.830, 0.830, 0.832, and 0.833, respectively, using the single-cluster update algorithm [6]. After thermalization we recorded 1 000 000 measurements (taken after 1, 1, 1, 1, 2, 4 clusters had been flipped) of the energy $E$ and the magnetization $M = (q \max\{n_i\} - V)/(q-1)$ in a time-series file, where $n_i \leq V$ denotes the number of spins of "orientation" $i = 1, \ldots, q$ in one lattice configuration. Obviously it is sufficient to store the integers $V/q \leq \max\{n_i\} \leq V$. From this data it is straightforward to compute all quantities of interest as a function of temperature by standard reweighting procedures [7]. The corresponding quantities per site are denoted in the following by $e = E/V$ and $m = M/V$.

More precisely we used reweighting to compute, e.g., the specific heat $C^{(i)}(\beta) = \beta^2 V \left(\langle e^2 \rangle - \langle e \rangle^2\right)$ for each replica labeled by the superindex $(i)$, and then performed the replica average $C(\beta) = [C^{(i)}(\beta)] \equiv (1/20) \sum_i^{20} C^{(i)}(\beta)$, denoted by the square brackets. To perform the replica average at the level of the $C^{(i)}$ (and *not* at the level of energy moments) is motivated by the general rule that quenched averages should be performed at the level of the free energy and not the partition function [8]. Finally, we determined the maximum, $C_{\max} = C(\beta_{C_{\max}})$, for each lattice size and studied the finite-size scaling (FSS) behaviour of $C_{\max}$ and $\beta_{C_{\max}}$. The error bars on the two quantities entering the FSS analysis are estimated by jack-kniving [9] over the 20 replicas. This takes into account the statistical errors on the estimates of each $C^{(i)}(\beta)$ as well as the fluctuations among the different $C^{(i)}(\beta)$ caused by the quenched randomness. The analysis of the magnetic susceptibility, $\chi(\beta) = \beta V \left([\langle m^2 \rangle - \langle m \rangle^2]\right)$ proceeds exactly along the same lines, yielding $\chi_{\max}$ and $\beta_{\chi_{\max}}$.

In the case of the (energetic) Binder parameter, usually defined on regular lattices as $B(\beta) = 1 - \langle e^4 \rangle / 3 \langle e^2 \rangle^2$, the proper definition of the replica average is less clear to us. In order to study this problem we have therefore computed the following three definitions which differ only by the replica averaging procedure: $B_1(\beta) = 1 - [\langle e^4 \rangle / 3 \langle e^2 \rangle^2]$, $B_2(\beta) = 1 - [\langle e^4 \rangle] / 3 [\langle e^2 \rangle^2]$, and $B_3(\beta) = 1 - [\langle e^4 \rangle] / 3 [\langle e^2 \rangle]^2$. While in spin glass simulations [10] usually the analogue of $B_3$ (with $e$ replaced by the overlap) is used, for a random bond Ising chain [11] a better scaling behaviour was observed for the analogue of $B_1$ (with $e$ replaced by $m$).

## 3 Results

Already a first qualitative inspection of our data gave a clear indication that the first-order nature of the phase transition on regular lattices (square, triangular, ...) persists on quenched random lattices. To make this statement more quantitative let us first consider the FSS of the specific-heat and susceptibility maxima. If the hypothesis of a first-order phase transition is correct, we expect for large system sizes an asymptotic FSS behaviour of the form [12–14]

$$C_{\max} = a_C + b_C V + \dots, \tag{2}$$

and

$$\chi_{\max} = a_\chi + b_\chi V + \dots. \tag{3}$$

Our data shown in Fig. 1 are clearly consistent with the Ansatz (2), (3). The least-square fits yield $a_C = 23.3(2.0), b_C = 0.0659(30)$, with a goodness-of-fit parameter $Q = 0.16$ (corresponding to a chi-square per degree of freedom of 1.7), and $a_\chi = -0.70(43), b_\chi = 0.0629(13)$, with $Q = 0.45$.

In Fig. 2 we show the scaling of the Binder parameter minima which is expected to be of the form

$$B_{i,\min} = a_{B_i} + b_{B_i}/V + \dots. \tag{4}$$

Since the data for $B_1$ and $B_2$ are almost indistinguishable, we have only shown $B_1$. Again the data confirms the hypothesis of a first-order phase transition, and from the fits we obtain $a_{B_1} = 0.6240(20)$, $b_{B_1} = -18.8(1.4)$, $Q = 0.17$, $a_{B_2} = 0.6236(22)$, $b_{B_1} = -18.5(1.4)$, $Q = 0.47$, and $a_{B_3} = 0.61125(68)$, $b_{B_3} = -16.45(71)$, $Q = 0.55$. Notice the much higher accuracy of $B_3$.

The locations of the extrema of $C(\beta)$, $\chi(\beta)$, and $B_i(\beta)$ define pseudo-transition points which, at a first-order phase transition, should scale according to

$$\beta_{C_{\max}} = \beta_0 + c_C/V + \dots, \tag{5}$$

etc., where $\beta_0$ is the infinite volume transition point. Our data and the corresponding fits through all data points are shown in Fig. 3. The resulting estimates for $\beta_0$ are 0.83360(14) from $C_{\max}$ ($Q = 0.51$), 0.83365(14) from $\chi_{\max}$ ($Q = 0.47$), and 0.83362(13) from $B_{3,\min}$ ($Q = 0.23$). On the scale of Fig. 3 the data points for $B_{1,\min}$ and $B_{2,\min}$ could hardly be disentangled from $B_{3,\min}$ and are therefore omitted. The results for $\beta_0$ are 0.83371(14) from $B_{1,\mathrm{min}}$ ($Q = 0.40$) and 0.83350(13) from $B_{2,\mathrm{min}}$ ($Q = 0.25$). Taking the average of the different estimates we finally obtain

$$\beta_0 = 0.83362 \pm 0.00013. \quad (6)$$

Notice that this value is very close to the exactly known transition point of the 8-state Potts model on a triangular lattice ($\beta_0^{\mathrm{triang.}} = 0.85666\ldots$) [4].

Finally we had a closer look at the replica fluctuations. In Fig. 4(a) we show the curves $C^{(i)}(\beta)$ for all 20 replica as well as the resulting replica average $C(\beta) = [C^{(i)}(\beta)]$. We see that all curves look very similar but are displaced by a constant amount in $\beta$. This is illustrated in Fig. 4(b) where we plot the same data vs $\beta - \beta_{\mathrm{max}}^{(i)}$, where $\beta_{\mathrm{max}}^{(i)}$ is the location of the specific-heat maximum for the i'th replica. Here all curves fall almost on top of each other and define quite nicely a kind of master curve. This suggests that the main effect of the randomness in the coordination numbers is to define a temperature reference point that is fluctuating among the different replica. This makes it plausible that coordination number randomness is not sufficiently strong to change the nature of the phase transition.

# 4 Conclusions

Summarizing, we have obtained clear numerical evidence for a first-order phase transition in the 8-state Potts model on quenched random lattices of Voronoi/Delaunay type. We can savely exclude the possibility of a cross-over to a continuous transition as was observed for a certain type of quenched bond randomness on square lattices [3].

In this brief note we have confined ourselves to FSS analyses of standard observables. It would be interesting to extend the analysis to quantities that are directly related to the probability distributions of the energy or magnetization, such as the interface tension and the "ratio-of-weight" definition of pseudo-transition points (which gave very precise estimates for regular square lattices) [14]. A quite elaborate study in this direction based on a much larger set of 128 replica will be published elsewhere [15].

# Acknowledgements

W.J. would like to thank D.P. Landau for useful discussions and communicating the results of Ref.[3] prior to publication. W.J. and R.V. were supported in part by NATO grant CRG940135. W.J. thanks the Deutsche Forschungsgemeinschaft for a Heisenberg fellowship, and also acknowledges support in part by EC grant ERBCHRXCT930343. The Monte Carlo simulations were performed on clusters of fast RISC workstations in Barcelona and Mainz.

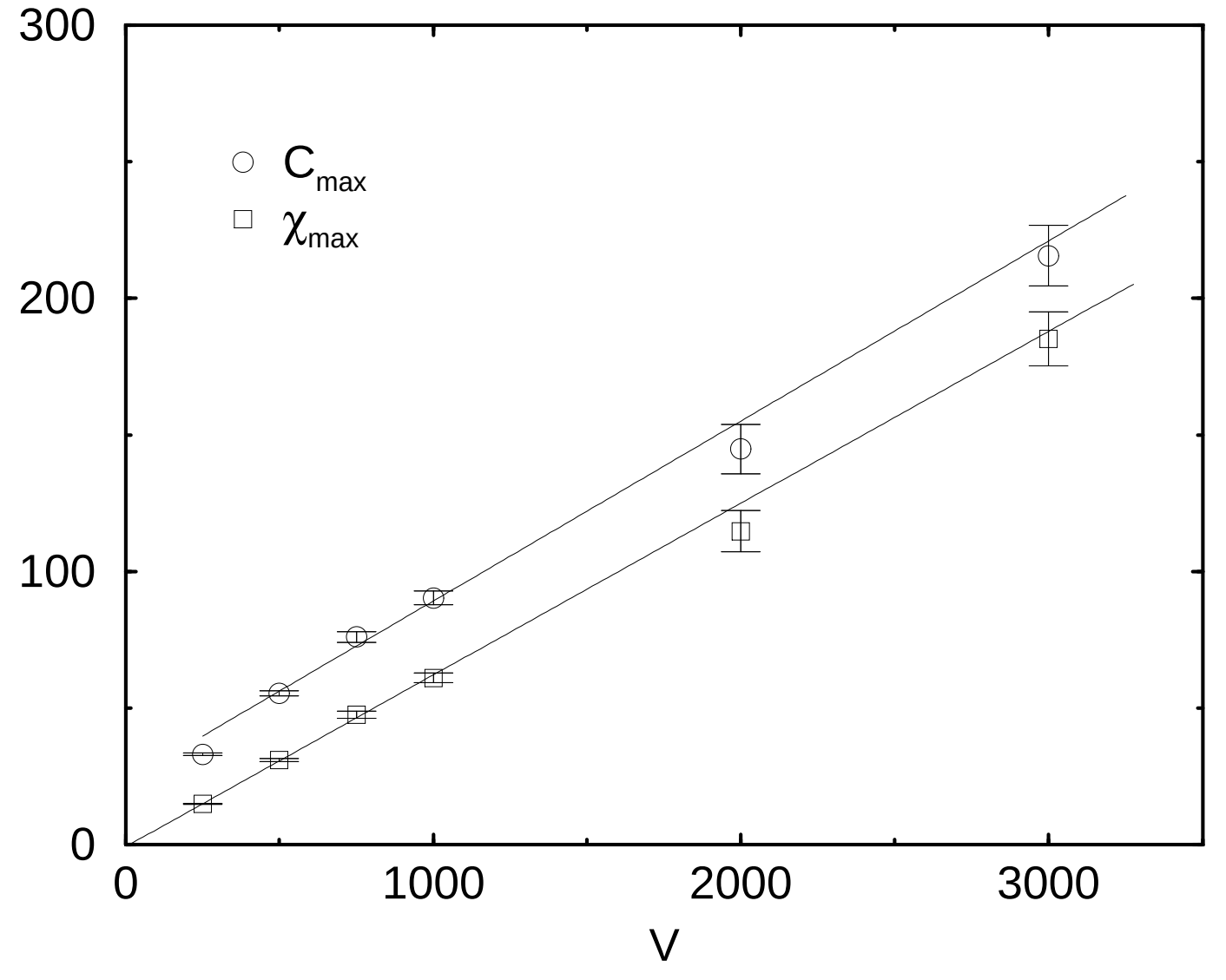

Fig. 1

Figure 1: *Finite-size scaling of specific-heat and susceptibility maxima.*

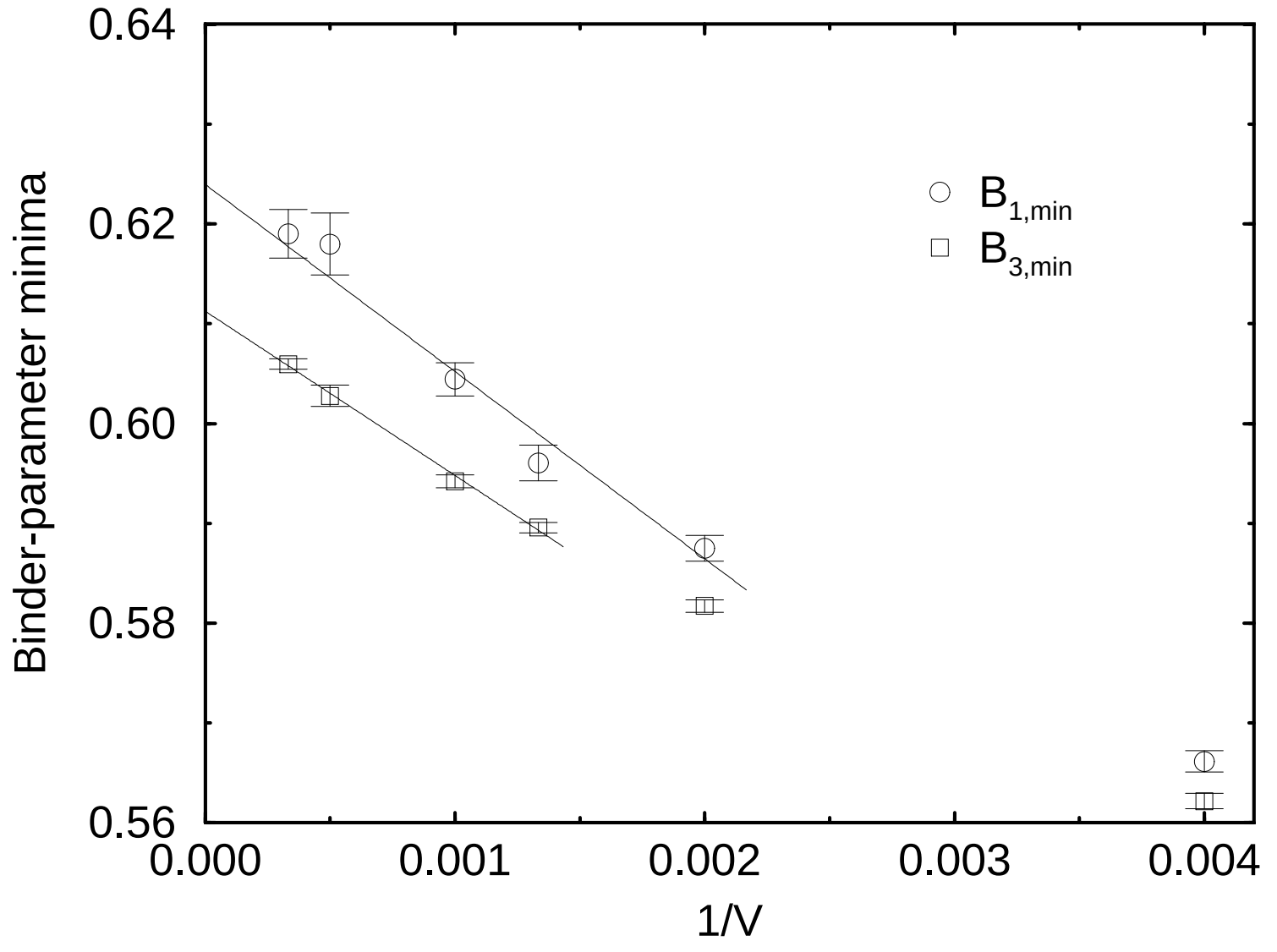

Fig. 2

Figure 2: *Finite-size scaling of Binder-parameter minima.*

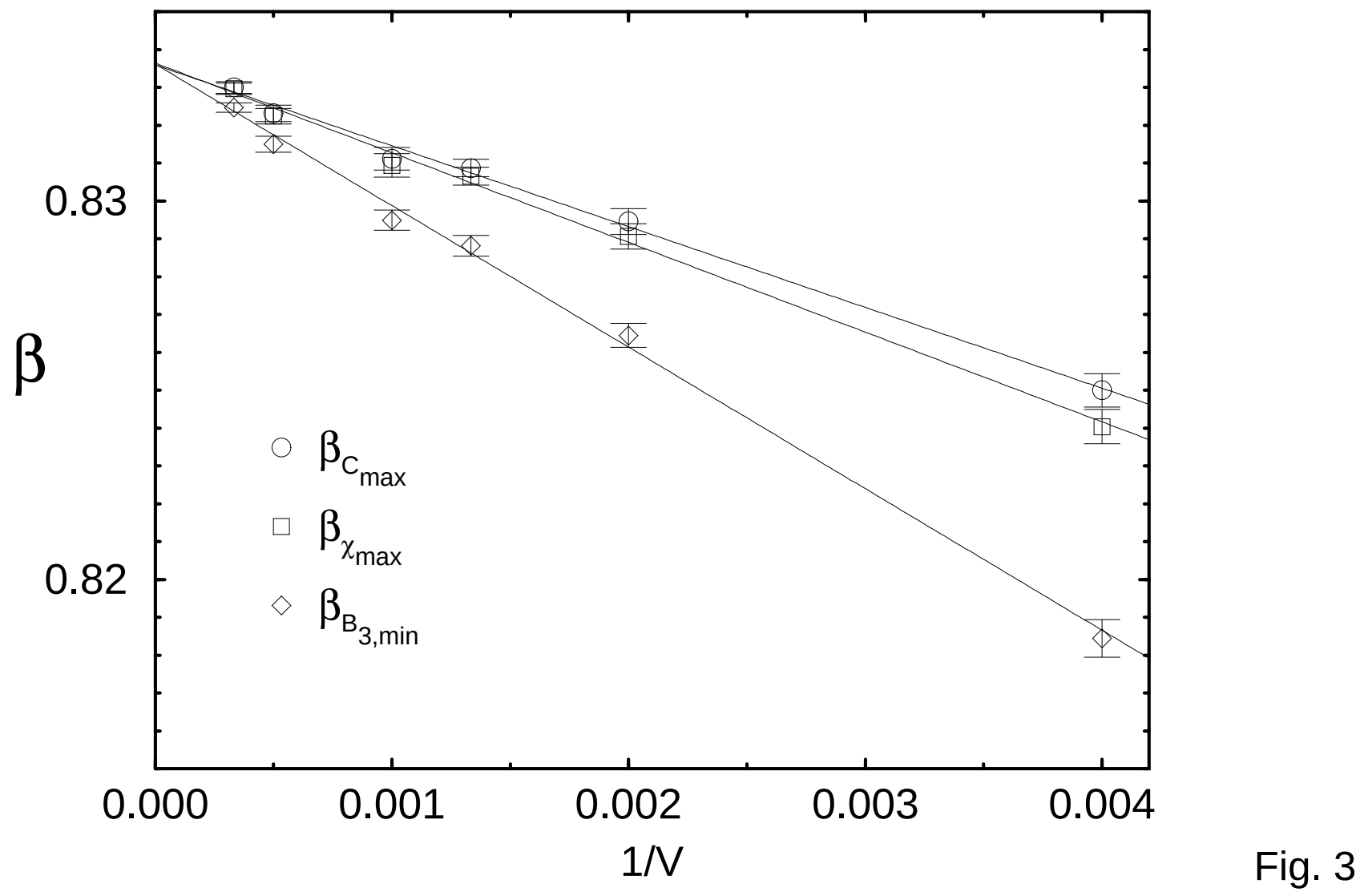

Figure 3: *Finite-size scaling of pseudo-transition points.*

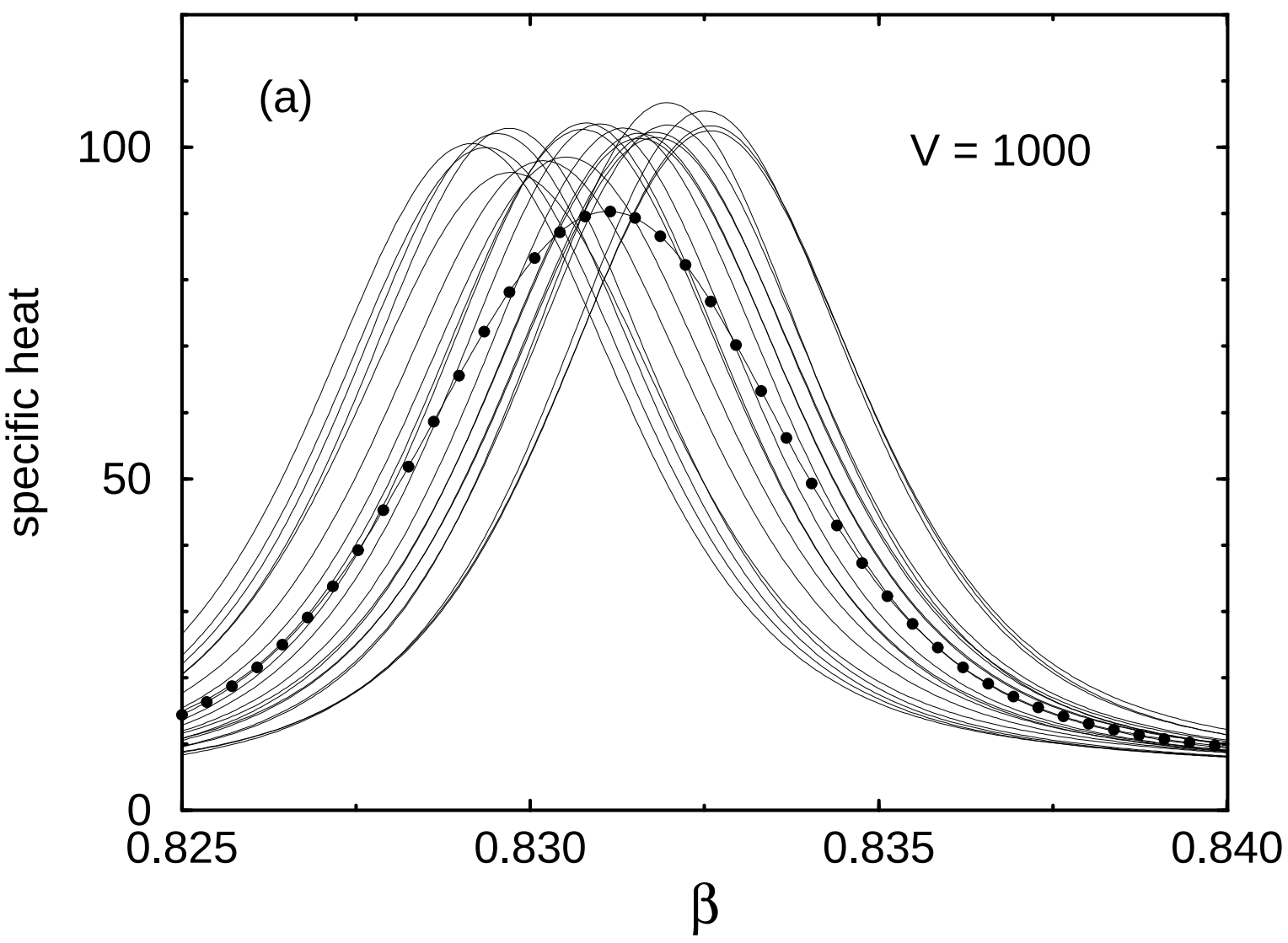

Fig. 4a

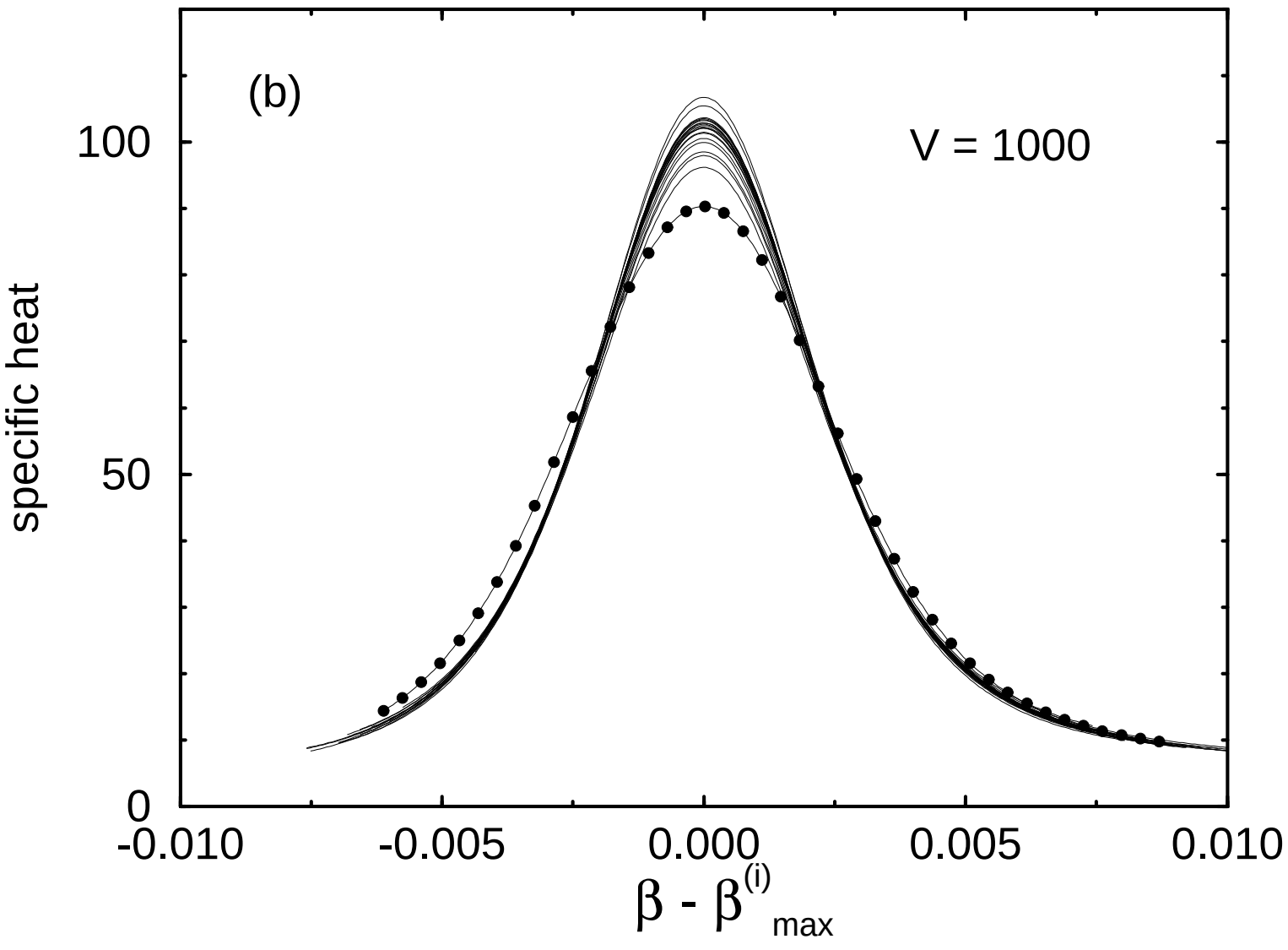

Fig. 4b

Figure 4: *(a) The specific heat $C^{(i)}$ for each of the 20 replica as a function of inverse temperature, and the replica average $C = [C^{(i)}]$ (marked by filled circles).*
*(b) The same data plotted vs $\beta - \beta^{(i)}_{\max}$, where $\beta^{(i)}_{\max}$ denotes the maximum location for the i'th replica. This shows that the main effect of the randomness in the coordination numbers can be parametrized by a random temperature offset.*